\documentclass{aa}  
 
\usepackage{graphicx}

\usepackage{amsmath}

\usepackage{txfonts}

\begin{document}

\headnote{Research Note}

\title{NGC 4262: a Virgo galaxy with an extended ultraviolet ring}

   \author{D. Bettoni
   \inst{1} 
    L.M.~Buson
   \inst{1} 
   \and 
   G. Galletta\inst{2}
    }

   \offprints{D. Bettoni}

   \institute{
  INAF Osservatorio Astronomico di Padova, vicolo  dell'Osservatorio~5, I-35122 Padova, Italy\\
  \email{daniela.bettoni@oapd.inaf.it,lucio.buson@oapd.inaf.it}\\
   \and
    Dipartimento di Astronomia, vicolo dell'Osservatorio~2, I-35122 Padova, Italy   \\
   \email{giuseppe.galletta@unipd.it} }
   \date{Received ... ; accepted .... 2010}

   \titlerunning{A galaxy with an extended ring}
   \authorrunning{Bettoni et al.}

  \abstract
  % context heading (optional)
    {The Galaxy Ultraviolet Explorer (GALEX) satellite has recently shown the presence of an extended, outer
     ring studded with UV-bright knots surrounding the lenticular galaxy NGC 4262. Such a structure---not detected
     in the optical---is coupled with a ring of atomic (HI) gas.}
% aims heading (mandatory)
    {We want to show that both star-forming and HI rings surrounding this SB0 galaxy share the same 
    radial distance from the galaxy center and spatial orientation. We want also to model the kinematics of the 
    ring(s) and of the galaxy body.}
     % methods heading (mandatory)
    {We make use of archive FUV and NUV GALEX data plus HI observations from the literature.}
    % results heading (mandatory)
    {We confirm that the UV-bright and atomic gas rings of NGC 4262 have the same extent and projected
    spatial orientation. Their kinematics is not coupled with that of the galaxy stars.}
% conclusions heading (optional), leave it empty if necessary 
   {It is possible that NGC 4262 has undergone a major gas stripping event in the past which gave origin to the present
   "necklace" of UV-bright knots. }

  \keywords{galaxies: general --- galaxies: elliptical and lenticular, cD---
   galaxies: individual: NGC 4262 --- galaxies: interactions}

   \maketitle

\section{Introduction}
Before the advent of the Galaxy Evolution Explorer (GALEX) UV-bright galaxy features  
without counterpart in the optical were unavoidably neglected. In particular GALEX has shown 
that over 30\% of spiral galaxies possess previously unknown UV-bright extensions of their 
optical disks (e.g. Thilker et al. 2007). This is also the case of the lenticular galaxy NGC~4262 in Virgo 
discussed here and surrounded by an extended, UV-emitting ring lying well beyond its optical image.

\begin{figure}
   \centering
   \includegraphics[width=8.5cm,height=8.5cm]{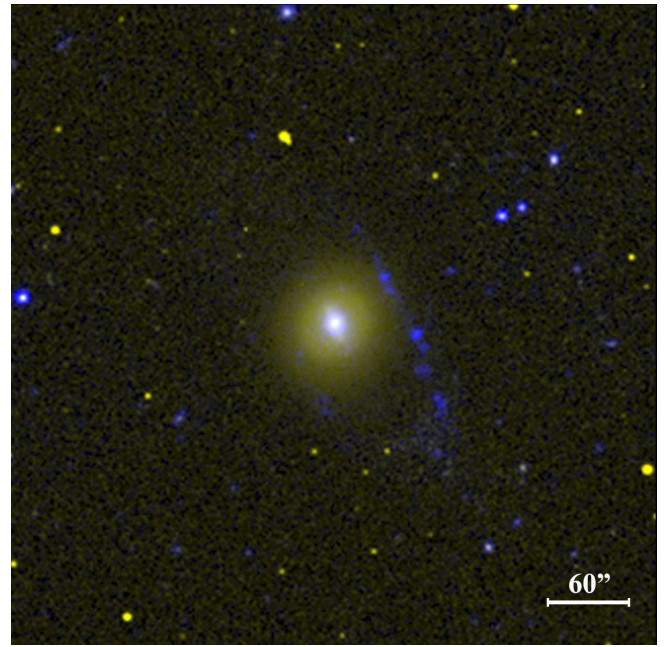}
 \caption{GALEX combined FUV and NUV image of NGC~4262 (blue=FUV, yellow=NUV). Note how individual knots 
 forming the far-UV-bright ring show up in this picture.}
\label{fig:iue_fos0}
\end{figure}

As far as NGC 4262 is concerned, Hubble Space Telescope Advanced Camera for Surveys (HST/ACS) imaging
(C{\^o}t{\'e} et al. 2006) unequivocally shows its morphology being typical of a SB0 at optical wavelengths. More precisely,
the galaxy has a bright, nearly spherical bulge, a bar, and a lens extending well outside the bar radius (Laurikainen et al. 
2010).
 
GALEX NUV (1771-2831 \AA) and FUV (1344-1786 \AA) images (Fig.~1) show instead an external knotty feature, 
something like "beads on a string"  likely made of individual, hot star clusters. In this respect, NGC 4262---having clustered 
UV-bright sources in its outer parts---could be classified as a Type~1 extended ultraviolet disk (XUV), a kind of feature discussed in detail by Thilker et al. (2007) and properly modeled by Bush et al. (2010); note, however,  that the lack of H$_\alpha$ observations does not allow us to state for certain we are looking at typical HII regions.

In this paper we want to put the question whether the (strict) coincidence of the UV knots and HI ring does reflect a past interaction event. The origin of the UV-bright ring is of necessity controversial: 
it could have formed together with the galaxy
itself, especially taking account that {\em tidal} stellar and gas arms are often offset (e.g. Mihos 2001) . On the other side, though the projected image rules out that NGC 4262 is a classical polar ring galaxy ( i.e. the ring is not at 90 degrees to the galaxy) and the knotting of the ring is similar to the knotting instability seen in the not-polar-ring Cartwheel galaxy, the presence of a ring surrounding a SB0 reminds shell and polar ring early-type galaxies, commonly thought to have undergone accretion and/or merging phenomena (e.g. Marino et al. 2009).

To our goal 
we want (i) to show that both star-forming and HI rings surrounding this SB0 galaxy (see Fig.~2) share the same radial distance from the galaxy center and spatial orientation and (ii) to model the kinematics of the ring(s) and of the galaxy body taking advantage of the existing innermost ionized gas velocity field (Sarzi et al. 2006) and the outer neutral hydrogen rotation curve (Krumm et al. 1985).

\section{Data and data reduction}

We de-archived both GALEX FUV (1344-1786~\AA) and NUV (1771-2831~\AA) images of NGC~4262 from deep frames (exposure time = 9.1 ksec [NUV], 7.2 ksec, [FUV], respectively) belonging to the program GI2 017 (P.I. Zhong Wang). 
The telescope has a very wide field of view (1.25 degrees diameter) and a spatial resolution $\sim$4".2 and $\sim$5".3 FWHM 
in FUV and NUV respectively, sampled with 1.5"$\times$1.5" pixels (Morrissey et al. 2007). 

We used FUV and NUV background-subtracted intensity images from the GALEX pipeline to compute
the ultraviolet integrated photometry and light profile of the galaxy. Background counts were estimated from the sky background image and high resolution relative response map provided by the GALEX pipeline itself. 
In addition we used optical Sloan Digital Sky survey (SDSS) archival images (Adekman-McCarthy et al. 2008) in the g (3630-58830 \AA) and i  (6340-8630 \AA) wavebands to obtain optical surface brightness profiles and (g-i) color profiles. 

Both UV and optical surface photometry have been derived 
by means of the IRAF\footnote{IRAF is written and supported by the IRAF programming group at the National Optical Astronomy Observatories (NOAO). NOAO is operated by the  Association of Universities for Research in Astronomy (AURA), Inc. under cooperative agreement with the National Science Foundation.}
STSDAS ELLIPSE routine. ELLIPSE computes a Fourier expansion for each successive isophote (Jedrzejewski 1987), 
producing radial luminosity and geometrical profiles, sampled in nested ellipses, 2 arcsecs wide.
These profiles are plotted versus galactocentric distance along the semi-major axis in Figs.~3 and 4.

In order to estimate the errors on UV (AB) magnitudes, we propagated the Poisson statistical errors on source and background counts. In addition to the statistical error, we added an uncertainty to account for systematic inaccuracies in the zero-point of the absolute calibration of 0.05 and 0.03 mag for FUV and NUV respectively (Morrissey et al. 2007). Our measured NUV total magnitude of NGC~4262 is consistent within errors with that measured by Gil de Paz et al. (2007). 
We derived also the FUV and NUV integrated fluxes of each  individual UV-bright 
knot in order to estimate its current star formation rate (SFR). They were computed as m(AB)$_{UV}$ = -2.5 $\times$
log CR$_{UV}$ + ZP where CR is the dead-time corrected, flat fielded count rate, and the zero points ZP=18.82 and 
ZP=20.08 in FUV and NUV respectively (Morrissey et al. 2007). We corrected both UV fluxes and luminosities by 
adopting the foreground reddening given by Schleghel et al. (1998), namely E(B-V)=0.035. With reference to the
extinction curve of Savage \& Mathis (1979), this translates into the extinction values 
A$_{FUV}$=0.28  and  A$_{NUV}$=0.32  respectively having adopted $\lambda$$_{eff}$=1516~\AA~ and 2267~\AA~ 
for the two above filters. 
\begin{figure*}
   \centering
   \includegraphics[width=14cm,height=14cm]{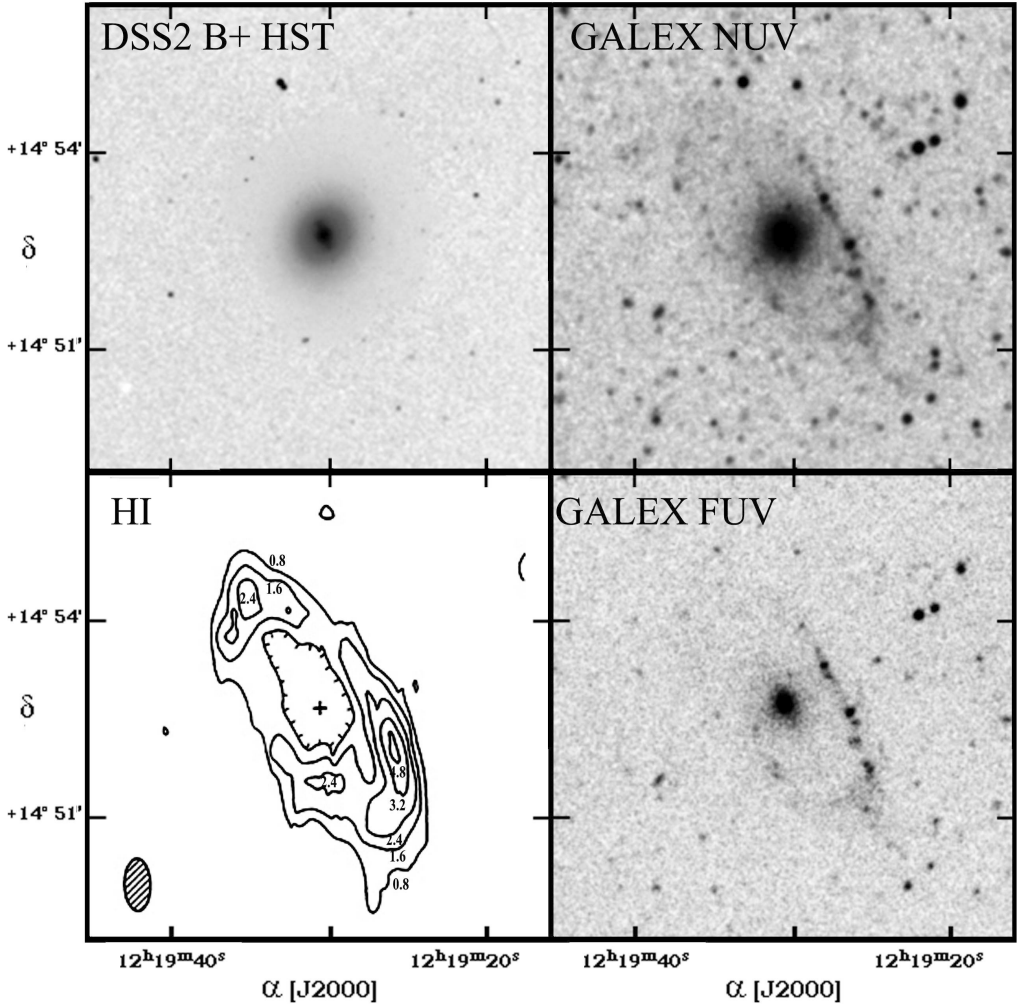}
 \caption{Comparison of optical (B+HST/ACS F475W) image of NGC 4262  (upper left),
 with GALEX NUV and FUV images (upper and lower right, respectively, and the smoothed
 HI column density (lower left). Contours are 0.8, 1..6, 2.4, 3.2, and 4.8 $\times$ 10$^{20}$ atoms cm$^{-2}$.
 The field of view of each panel is $\sim$5'.3$\times$5'.3. The ring physical size turns out to be $\sim$8.5~kpc while the galaxy body---as seen in the NUV---has a radius of only $\sim$4~kpc.}
\label{fig:iue_fos}
\end{figure*}

\section{Galaxy 
and UV-bright clumps photometry}

The UV and optical surface brightness and color profiles of the galaxy body are shown in Fig.~3.  
The luminosity profiles of NGC 4262 at various wavebands indicate the predominance of a peaked distribution, that appears linear when plotted in r$^{1/4}$,  inside $\sim$16 arcsec and of a exponential distribution outside this radius. So, data was fitted with a double-component curve made by a sum of a r$^{1/4}$ law and an exponential law, spanning all the range of observed luminosity profiles. A preliminary fit was performed with a single component only, inside and outside 16 arcsec, to define the starting parameters, that are the effective radius r$_e$ and the apparent surface brightness  at the same distance $\mu_e$. Then a model with both components was run, by adjusting the  r$_e$ and  $\mu_e$ parameters for both luminosity laws with the goal of minimize the residuals between observed and calculated surface brightness. 
To perform the fit the very inner region of the profiles, affected by the seeing (PSF= 2" in the optical and PSF= 5" in the UV), were excluded. The best fitting parameters are listed in Table~1. 

Then, the total magnitudes of both components, bulge and disc  were calculated and the resulting total galaxy magnitude for every waveband was obtained. These values also are listed in Table~1. We found a very good agreement with literature values, within a few hundredths of  magnitude in i, g and NUV values.  The error in effective radii and surface brightness were calculated from the errors in the fitting parameters (slope and intercept) and are also reported in the table. Errors in total magnitudes were obtained from the correlation coefficient of the model fit vs the observed values.

As one can see the disk contribution decreases as we go towards shorter wavelengths, and in the FUV profile is only barely visible, In the Fig.~3 an arrow marks the radial position where the disk starts to prevail. We note that---unlike optical color g-i---the FUV-NUV color becomes much bluer in the innermost region of the galaxy (the innermost UV color point represents the mean value inside the GALEX PSF).

In Fig.~4 we show also the ellipticity and position angle profiles, for all the four bands. The galaxy is seen almost face on, as the apparent disk ellipticity is  0.1 corresponding to an inclination of $\sim$26$^{\circ}$, assuming an intrinsic flattening of 0.25 for the disk. In the g and i images the bar signature is clearly visible from 5 to 20 arcsecs, with an increase of ellipticity 
(till $\epsilon_{bar}$=0.4) and a sharp change in the major axis to P.A=25$^{\circ}$. This is indeed the position angle of the major axis of the bar and is almost coincident with the P.A. of the major axis of the gas that reproduces its observed kinematics (both of the neutral and ionized components) as described in Section~5. In the NUV ellipticity and P.A. profiles the signature of the bar is still present in the same region, whereas almost disappears in the FUV profiles where only the bulge component is visible. We note that the bar in the NUV is rounder than in the optical ($\epsilon_{barUV}$=0.3). In Fig.~5 (upper panel) is shown a schematic view of the galaxy (the apparent flattening on the sky of the stellar disk (black circles), the bar (red ovals) and outer HI ring (green ovals), while in Fig.~5 (lower panel) the edge-on view of the system with the line of sight is reported. On the upper right of each figure are given the used geometrical parameters.

In addition to the luminosity profiles we measured the integrated m$_{AB}$ magnitudes and 
colors of the UV-emitting clumps that are given in Table~2. The brightest clumps we measured were selected by optical inspection and all the magnitudes are obtained inside an aperture of 7 arcsec in diameter. At the HST-based, adopted distance of 14.6 Mpc (Jord\'an et al. 2005) our (GALEX) spatial resolution is $\sim$400~pc and we cannot establish whether their physical size is in line with typical Galactic loose OB associations ($\sim$80-100 pc). In fact the clump measured FWHMs are comparable with the large GALEX PSF (4-6'') and, as such, they are not resolved.  As stated above, the adopted distance to convert AB magnitudes
to luminosities is 14.6~Mpc. A map of individual measured, UV-bright clumps along the ring 
is shown in Fig.~6 Such a figure shows the full properly smoothed HI column density map (Krumm et al 1985)
superposed to the recent GALEX image of the galaxy and its ring. The resulting match
between the UV and HI rings is outstanding and clearly represent the same feature. What is more, the regions of highest HI condensation correspond to the brightest complexes in the UV.

\section{Star formation rate}
The present-day star formation rate of each UV-bright blob along the ring can be derived---following 
Kennicut (1998)---using its UV continuum luminosity and the relation\\

 SFR$_{FUV}$ (M$_\odot$ yr$^{-1}$) = 1.4 $\times$ 10$^{-28}$L$_{FUV}$(ergs s$^{-1}$ Hz$^{-1}$).\\
 
 \noindent
FUV fluxes, luminosities and SFRs are given in Table ~2.
Note that our UV evidence of star formation within the galaxy body is only apparently in contrast with the lack of
star formation derived in the infrared by Shapiro et al. (2010). In fact their low ratio between 8.0~$\mu$m and 
3.6~$\mu$m SPITZER fluxes could be ascribed to a low amount of dust as well.

\begin{figure}
\includegraphics[width=9.0cm,height=8.0cm]{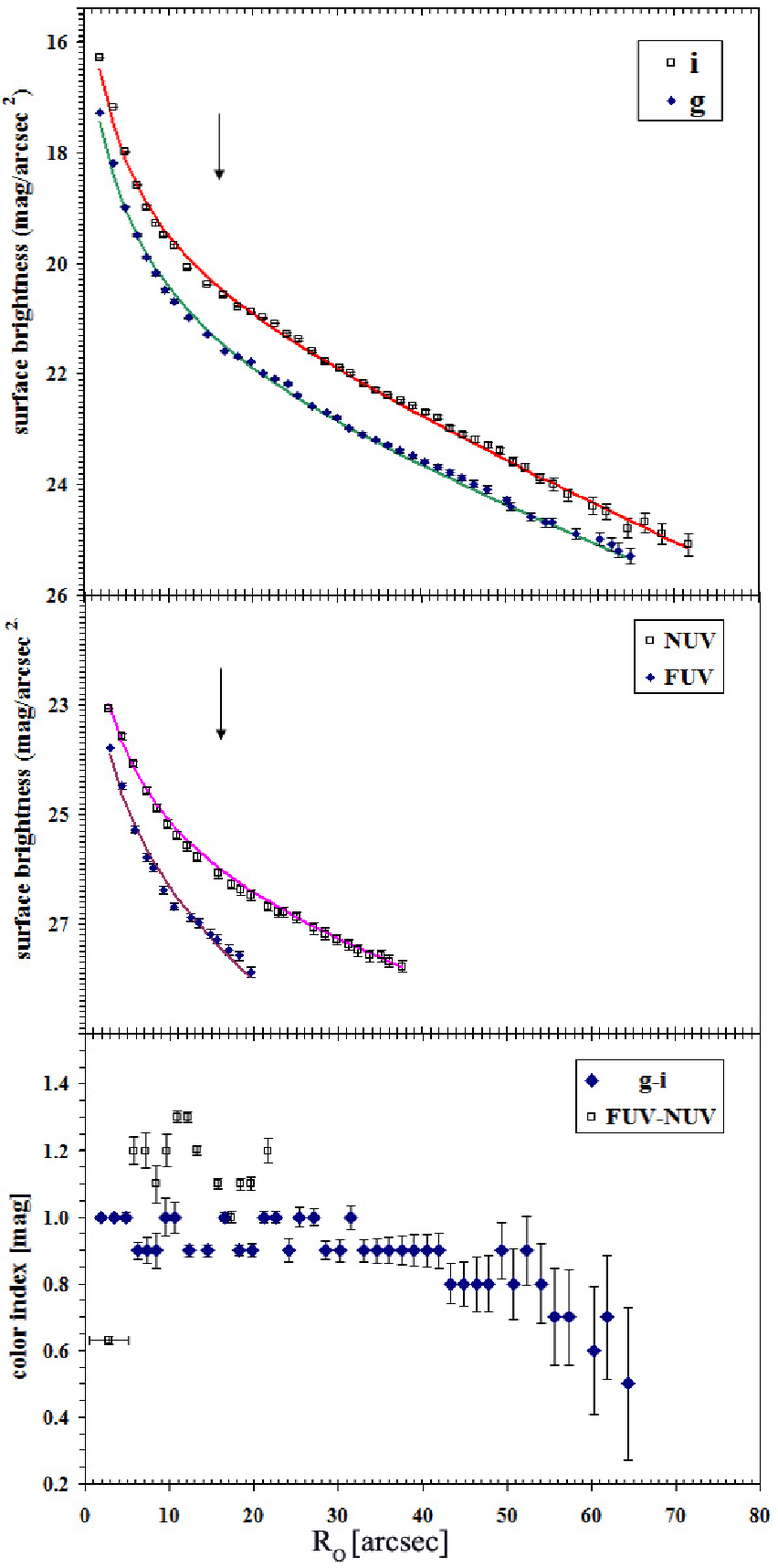}
\caption{UV/Optical surface brightness and color profiles of NGC 4262 main body. 
In each panel we always plot the sum of the two single fitted profiles (arrows mark the radial distance where the disk starts to prevail). Upper panel shows the optical (g,i) profiles, while mid panel gives the UV (NUV, FUV) corresponding profiles. As one can see the disk contribution decrease as we go towards shorter wavelengths, and in the FUV profile is only barely visible. Both optical (g-i) and UV (FUV-NUV) color profiles are presented in the bottom panel.  Note that the innermost UV color point represents the mean value inside the GALEX PSF.}
\label{fig:iue_fos4}
\end{figure}

\begin{figure}
\includegraphics[width=9.0cm,height=7.0cm]{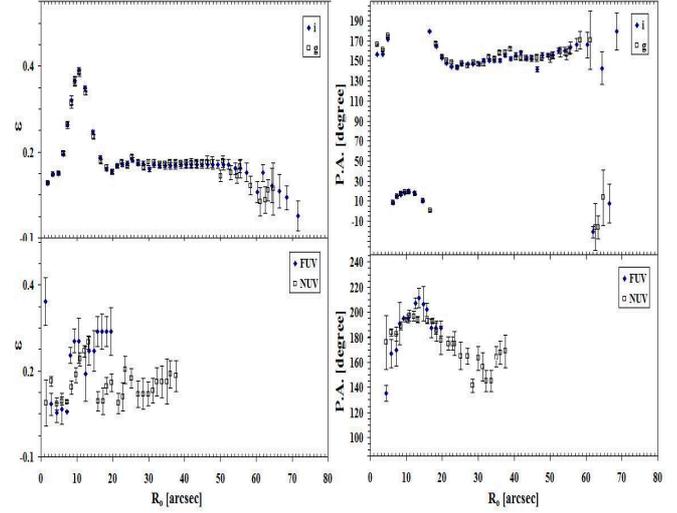}
\caption{Position angle and ellipticity profiles for both optical and UV images of NGC~4262.
Upper left panel: optical ellipticity profile. Upper right panel: P.A. optical profile. Note the remarkable
change both in ellipticity and P.A. corresponding to the optical galaxy bar. Lower left panel: UV ellipticity
profile. Lower right panel: P.A. UV profile. Note the disappearance of the bar's effect at ultraviolet wavelengths.}
\label{fig:iue_fos4}
\end{figure}

\begin{figure}
\centering
   \includegraphics[width=8.5cm,height=8.5cm]{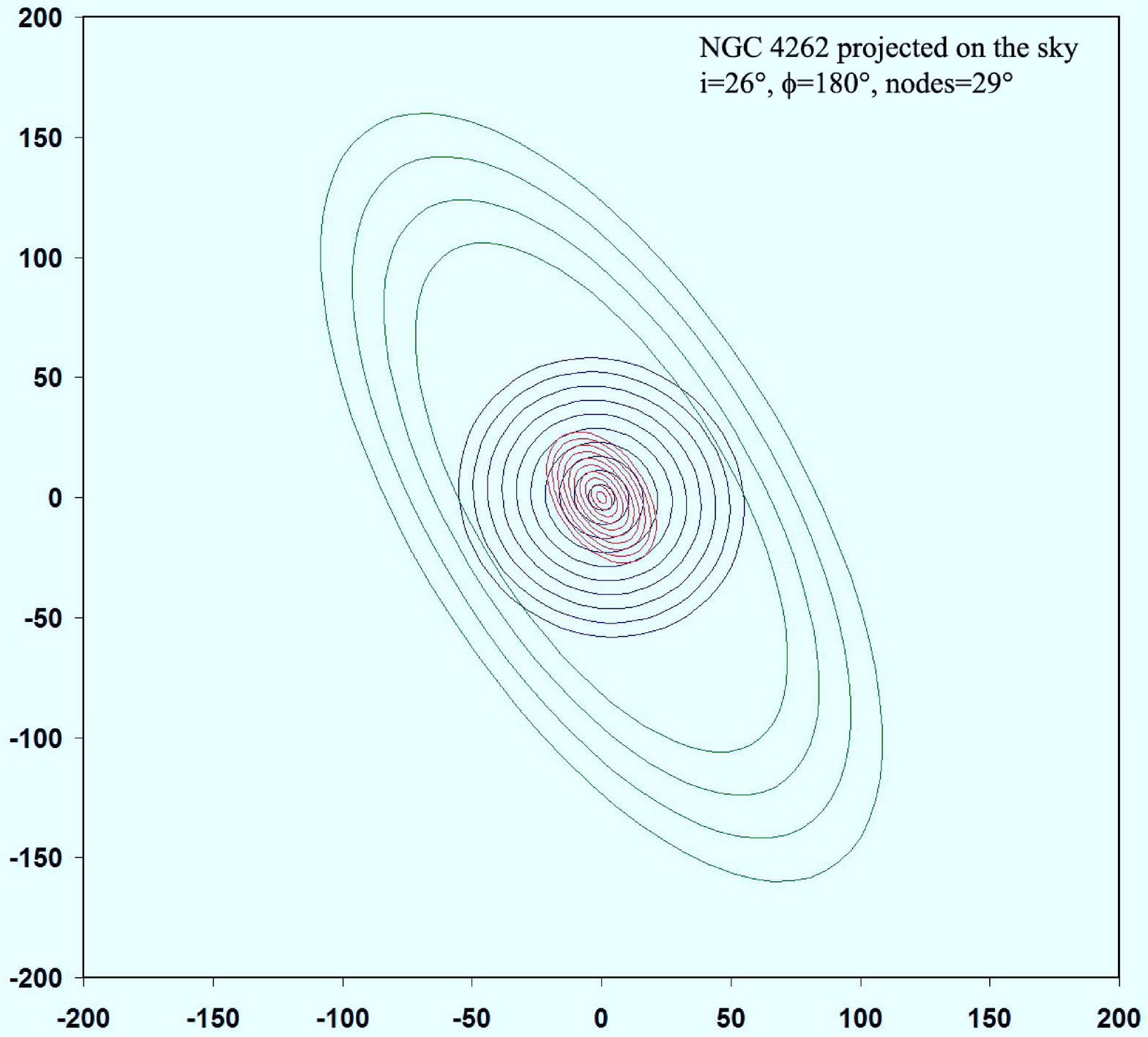}
 \includegraphics[width=8.5cm,height=8.5cm]{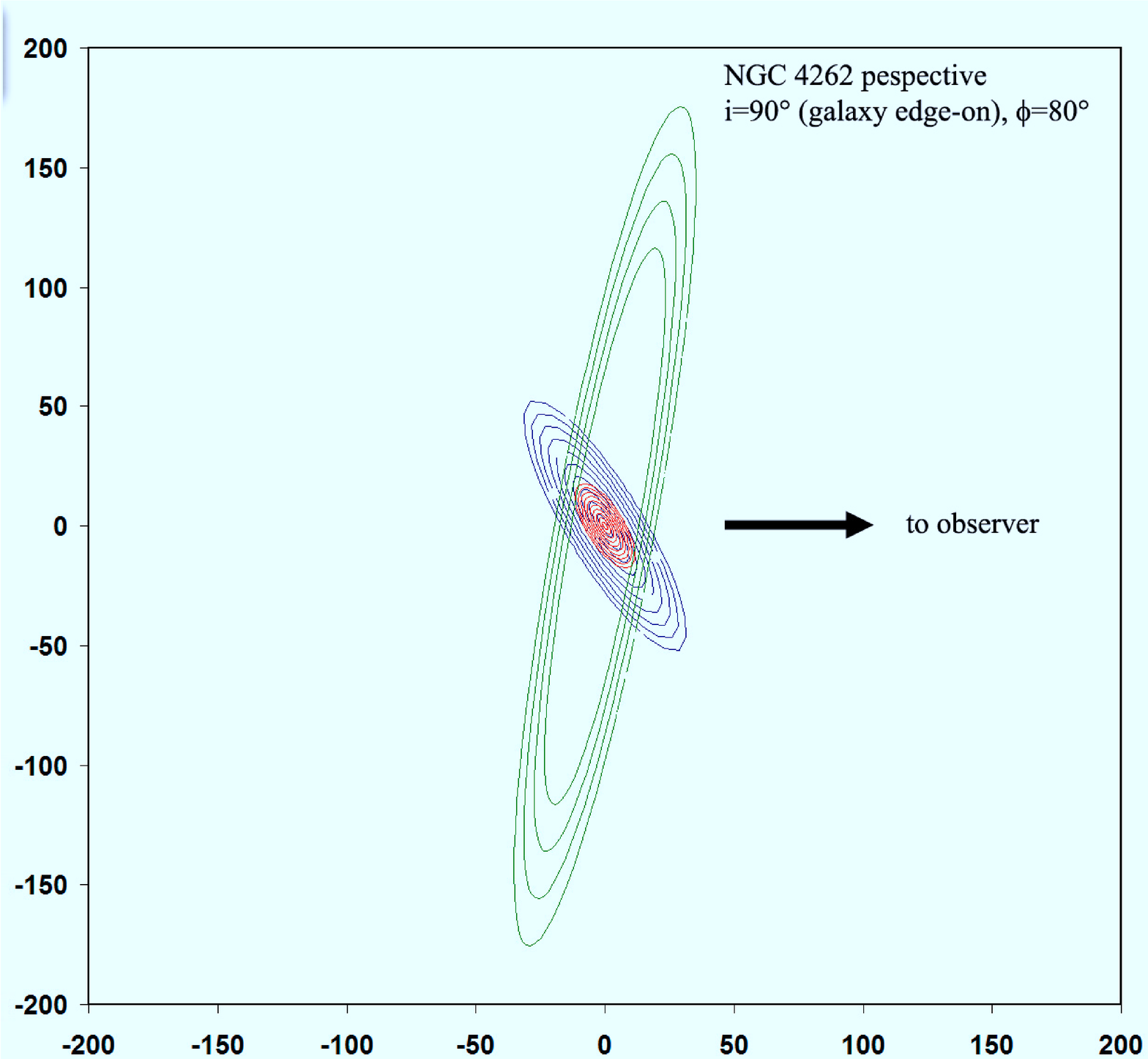}
  \caption{ A schematic view of the galaxy. Upper panel: the apparent flattening on the sky of the stellar disk (black circles), the bar (red ovals) and outer HI ring (green ovals) are shown. Lower panel: Edge-on view of the system with the line of sight reported. On upper right of the figures the used geometrical parameters are 
 reported.}
\label{fig:iue_fos6}
\end{figure}

% \begin{figure}
%\centering
  % \includegraphics[width=8.5cm,height=8.5cm]{Modello_N4262_2.eps}
 %\caption{ Edge-on view of the system with the line of sight reported. On upper right of the figure are reported the used geometrical parameters.}
%\label{fig:iue_fos7}
%\end{figure}

\begin{table*}
\caption{Photometric parameters from the fit of the luminosity profile}
\label{Photometric parameters}
\begin{tabular}{lccccc}
\hline\hline
Band. & $m_{tot}$  & $r_{e~(bulge)}$ & $\mu_{e~(bulge)}$ & $r_{e~(disk)}$ & $\mu_{e~(disk)}$\\
& mag   & arcsec & $mag/"^2$ & arcsec & $mag/"^2$\\
\hline
i & 10.96$\pm$0.01 & 5.2$\pm$0.92 & 18.40$\pm$0.40  & 22.80$\pm$0.22 & 21.82$\pm$0.03 \\
g & 11.95$\pm$0.01 & 6.0$\pm$0.80 & 19.53$\pm$0.29 & 26.81$\pm$0.36 & 23.33$\pm$0.04 \\
NUV & 16.26$\pm$0.02 & 7.99$\pm$0.99 & 24.93$\pm$0.26 & 31.98$\pm$1.5 & 27.96$\pm$0.08\\
FUV & 17.66$\pm$0.1 & 6.03$\pm$1.11 & 25.33$\pm$0.42 & 15.2$\pm$8.1 & 29.3$\pm$1.13 \\
\hline
\end{tabular}
\end{table*}

\begin{table*}[h!]
\caption{Integrated photometric properties}
\begin{tabular}{lcccccccccc}
\hline
Object & R.A. & DEC & {m$_{FUV}$(AB)} & {F$_{FUV}$} & {L$_{FUV}$}& {m$_{NUV}$(AB)} & {F$_{NUV}$} & {L$_{NUV}$} & {FUV-NUV} & {SFR$_{FUV}$}\\
 & (2000) & (2000) & &  & &  & & & & \\
\hline
\hline
Galaxy & 184.8776  & 14.87777 & 17.66$\pm$0.12 & 2.466 & 6.41 & 16.26$\pm$0.01 & 8.241 & 21.43 & 1.31 & 8.9\\
Clump 1 & 184.8732  & 14.89737 & 22.68$\pm$0.39 & 0.0308 & 0.0800 & 22.06$\pm$0.31 & 0.0544 & 0.142 & 0.62 & 0.112\\
Clump 2 & 184.8762  & 14.88732 & 21.18$\pm$0.24 & 0.123 & 0.318 & 20.92$\pm$0.19 & 0.156 & 0.405 & 0.26 & 0.446\\
Clump 3  & 184.8603 & 14.87558 & 20.69$\pm$0.19 & 0.192 & 0.500 & 20.59$\pm$0.18 & 0.212 & 0.548 & 0.10 & 0.700\\
Clump 4  & 184.8589 & 14.86833 & 21.61$\pm$0.31 & 0.082 & 0.214 & 21.23$\pm$0.24 & 0.117 & 0.304 & 0.38 & 0.300\\
Clump 5  & 184.8554 & 14.86131 & 21.32$\pm$0.22  & 0.108 & 0.280 & 21.15$\pm$0.23 & 0.123 & 0.327 & 0.17 & 0.392\\
Clump 6  & 184.8801 & 14.86162 & 22.13$\pm$0.31 & 0.0510 & 0.133 & 21.70$\pm$0.30 & 0.0759 & 0.197 &  0.43 & 0.186\\
\hline
\end{tabular}\\
\label{tab:fos_indb}
\small{Notes: (1) Coordinates are given in degrees\\ 
(2) Fluxes are in units of $10^{-27}$ erg s$^{-1}$cm $^{-2}$$Hz^{-1}$ \\
(3) Luminosities are in units of $10^{26}$
erg s$^{-1}$ $Hz^{-1}$\\ (4)  SFRs are in units of $10^{-2}$ M$_{\odot}$ yr$^{-1}$}
\end{table*}

\section{Galaxy and ring kinematics}
\begin{figure*}
\centering
   \includegraphics[width=14cm,height=14cm]{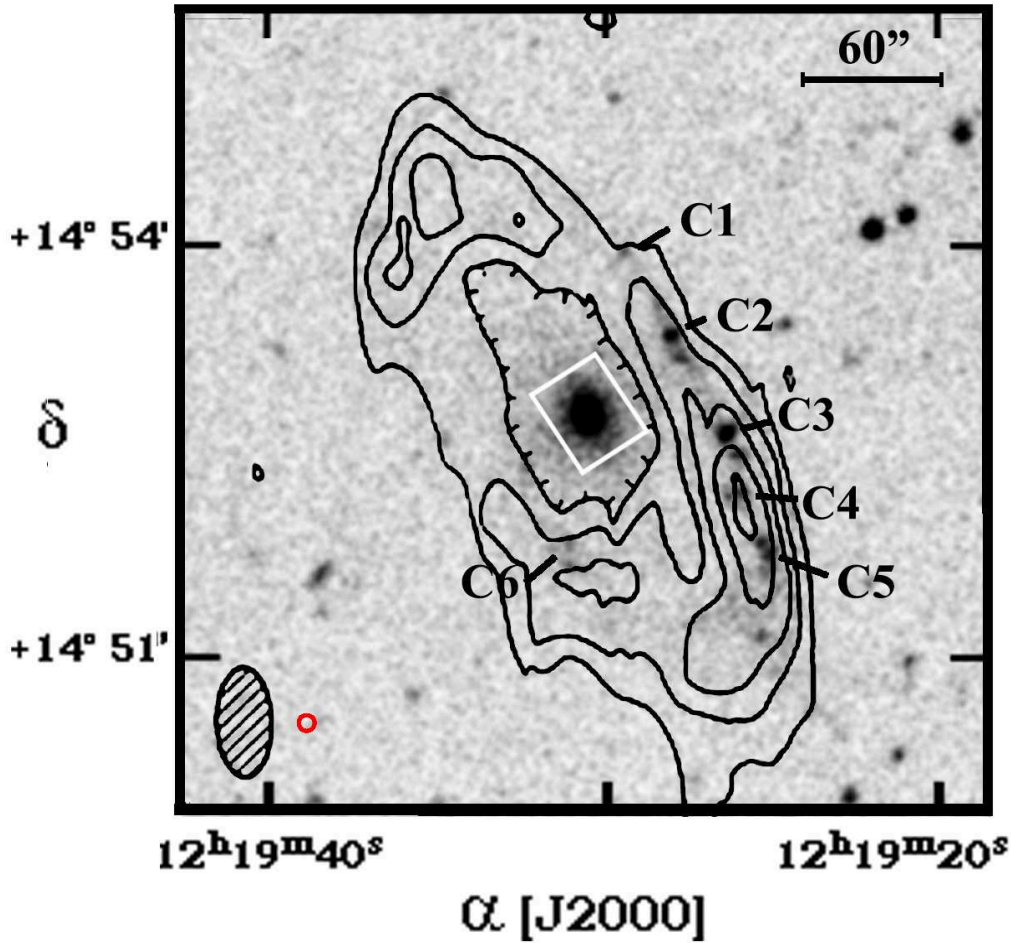}
 \caption{GALEX FUV  image of NGC 4262  with superimposed the labels of individual, photometered UV-bright  clumps 
 along the outer ring. Their photometric parameters
 are listed in Tab.~1.The field of view is $\sim$4'.3$\times$4'.3. The red empty circle shows the size of the photometric aperture used for all UV blobs (7 arcsec diameter), while the white empty square on the galaxy nucleus represents the size of the SAURON field of view (33" $\times$ 41").}
\label{fig:iue_fos5}
\end{figure*}
The innermost kinematics of NGC~4262 appears quite complex. Sarzi et al. (2006) write: "... this strongly
barred galaxy shows an integral-sign pattern in the gas distribution and a twisted gas velocity field. In addition this object 
shows peculiar asymmetric distributions for the central values of sigma and for OIII/H$\beta$ ratio. The gas and stellar
kinematics are strongly decoupled."

To reproduce the gas shape and kinematics we used a model composed by a set of rings (described in Appendix~A) inclined with respect to the sky plane and to the galaxy's reference plane. The intrinsic gas rotation curve is a typical Brandt (1960) curve with a steep increase in the first arcsecs from nucleus and an almost flat part outside (Fig.~7). The observed values for atomic gas distribution and kinematics are from HI observations by Krumm et al. (1985) while the ionised gas kinematics has been studied by Sarzi et al. (2006). We deduced the values from the bidimensional maps and graphs of Krumm et al. (1985)  and from a table of data used for the Sarzi et al. (2006) work and kindly sent us by dr. Marc Sarzi (the size of the SAURON
velocity field is overplotted as a white rectangle in Fig.~6). For optical emission lines error bars have been included as the r.m.s. of the velocities of each line. For HI points error bars have been taken 
from Krumm et al. (1985). The rotation curve has been extracted along a P.A. of 29$^\circ$, that corresponds to the sky elongation of the outer HI and UV ring, from 80$\arcsec$ to 120$\arcsec$. Then, a 100-ring model has been generated that fits the ring shape in the inner part (flattening and P.A.). This generates the $i, \phi$ geometrical parameters described in Appendix A. The maximum extent of the model (100-th ring) has been adjusted to 120$\arcsec$, the maximum HI ring extent. As a first approach, no warping or twisting of the set of rings has been applied (in the model $\Delta\delta_n$=0 and $\Delta\alpha_n$=0 for all the rings). We found, in agreement with Krumm et al. (1985), that the inclination of the set of rings is 65$^\circ$ and their line of the nodes is at P.A.=29$^\circ$. 

After a solution for geometrical parameter has been found, we tried different solutions to fit the observed rotation curve both of HI and ionised gas. As a result, we found that the maximum rotation of the ring holds at $\sim$13$\arcsec$ with a maximum velocity of $\sim$210 km/s $\pm$ 10 km/s (see Fig.~7). The inner kinematics of the ionised gas may be influenced by the bar, that appears elongated at the same P.A. of the ring. The peaks of velocity that appear in [O III]and H$\beta$ at few arcsec fron nucleus may be due to this deviation from circularity. Unfortunately, no observed data for galaxy motions between 20$\arcsec$ and 80$\arcsec$ exist, so we tried the simplest possible solution fitting inner and outer data. 

The intensity distribution of [O III] and H$\beta$ lines published by Sarzi et al. (2006) indicates a possible elongation of the gas to P.A.=0$^\circ$. If real, this may indicate a twisting or spinning in the space of the gas ring toward the center. As an alternative, they may be a surposition of the gas circulating inside the bar in elliptic streamings with that of the disc in circular motion. But the map of the ionised gas distribution appears quite irregular, so we preferred do not search solutions with warped discs, to avoid  increasing the number of variables present in the model.  

\begin{figure}
\centering
\includegraphics[width=7.5cm]{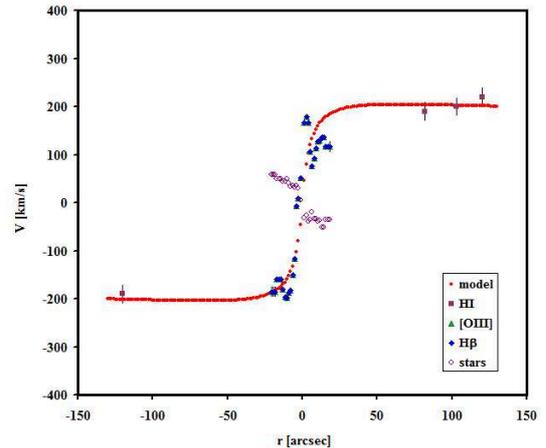} 
\caption{ The observed rotation curve of NGC 4262 extracted along the line of the nodes of the external ring (P.A.=29$^\circ$). Inner data are from Sarzi et al. (2006)  while full squares indicates the HI velocity deduced from Krumm et al. (1985). The model rotation curve crosses the points both in the inner and outer regions. Note the counterrotation gas-stars in the inner regions of NGC 4262.} 
\label{rotation}
\end{figure}

\section{Origin of the stellar ring}
Taking into account the clear symmetry of the HI ring around NGC 4262, it is unlikely it comes from a primordial
cloud of neutral hydrogen such that discussed in detail by Thilker et al. (2009). On the contrary, Bekki et al. (2005) do model the stripping of rings and arcs of cold gas as due to interaction with other galaxies. In this context, a stream of gas pulled out of the disk of the galaxy (or contributed by a perturber) forms stars when it is compressed.
Such an interaction scenario is supported also by a further hint, namely the observed inner decoupling of gas and stars velocity fields in NGC~4262 (Sarzi et al. 2006). In the specific case of NGC 4262 the existence of a mutual interaction with the other Virgo galaxy NGC 4254 has been proposed by Chyzy et al. (2007), while another example of such merger-induced, ringed 
lenticular galaxies could be NGC~404 (Thilker et al. 2010). As far as the UV-detected, disk's star formation is concerned,
is likely occurring independently of the above possible (ring-forming) interaction event.

\section{Summary and Conclusions}
Thanks to the UV-sensitive GALEX satellite we were able to detect an extended, UV-bright ring
surrounding the otherwise normal SB0 galaxy NGC~4262. Such a feature (not recognizable in the optical)
appears to host several knots, likely consisting of hot star clusters. In this respect, NGC 4262---having clustered UV-bright
sources in its outer parts---could be classified as a Type~1 extended ultraviolet disk (XUV). About the origin of such a structure, one should be aware that theoretical models (e.g. Bekki 2005; Higdon \& Higdon 2010) ascribe the onset of rings and arcs of cold gas as well as the formation of young star rings to the past interaction with other galaxies. As a consequence, taking account also of the observed inner decoupling of gas and stars velocity fields (Sarzi et al. 2006), we are pretty confident that a past major interaction episode underwent by NGC~4262 is responsible of the onset of the UV-bright ring we see today.

\begin{acknowledgements}
We are indebted with Marc Sarzi and SAURON Team for providing their kinematical data. We thank the anonymous
referee for his/her comments improving our paper.
\end{acknowledgements}

\appendix
\section{System geometry}\label{geometry}

\begin{figure}
\centering
\includegraphics[width=7.5cm]{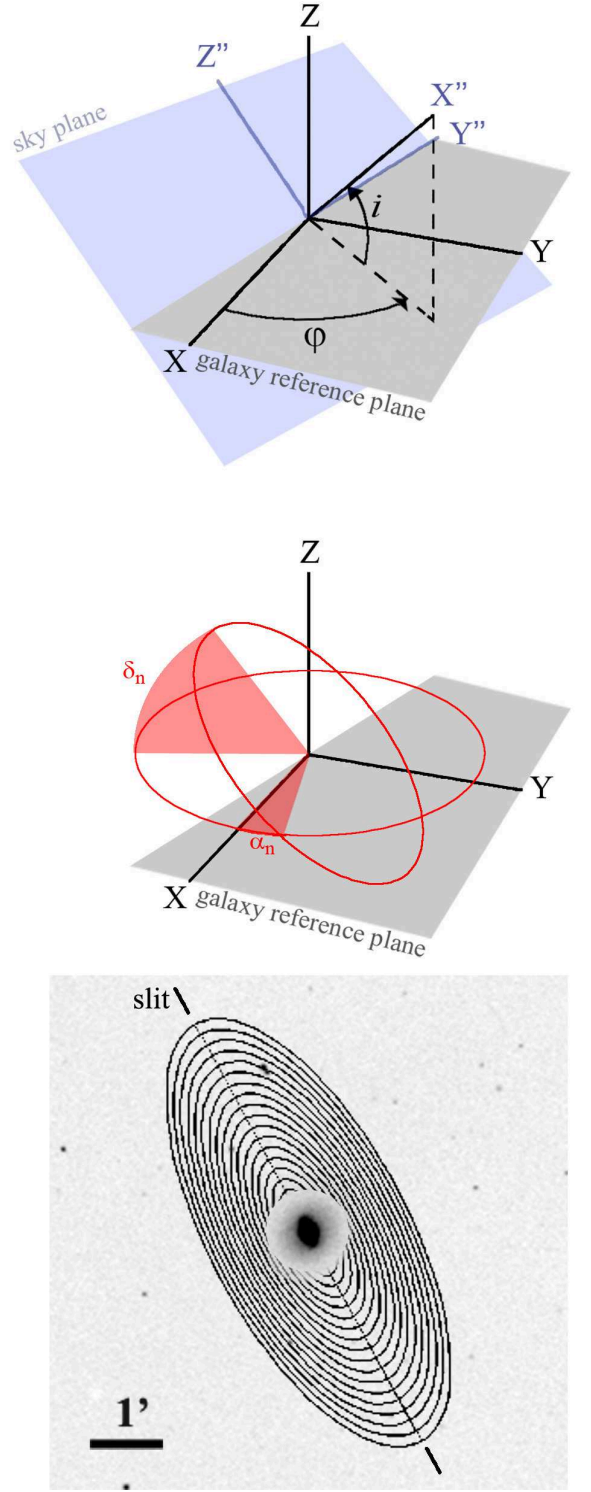} 
\caption{ {\it top panel:} The angles of the observer with respect to the galaxy reference plane. {\it{mid-panel:}} The inclination of a ring with respect to the galaxy reference plane. {\it{bottom panel:}} Image of the model rings plotted on a DSS I image of NGC 4262. A circular "hole" has been created in the rings image only for showing purpose, to make the galaxy visible. The outermost ring is at 200 arcsec, corresponding to the extension of the HI ring visible in the other figures of this paper. The straight line across the rings indicate the position of the slit used for the extraction of the model velocity values.}
\label{geom1}
\end{figure}

We define an {\it observer reference system} X''Y''Z'' whose Y''Z'' plane coincides with the sky plane and the X'' axis represents the line-of-sight. The HI distribution is assumed to lie on the XY plane of a {\it galaxy reference system} in a set of concentric, circular rings. The two sistems are inclined by the angles $i$ and $\phi$ (see Figure \ref{geom1}), that are the inclination and the azimuth of the line of sight  with respect to XYZ. The relation between the coordinates of the two systems is described by the rotation matrix R, as discussed in Galletta (1983):  

\begin{equation}
\begin{pmatrix}
 x'' \\ 
 y'' \\	
 z'' \\	
\end{pmatrix}
=
 \begin{pmatrix}
R_{11} & R_{12} & R_{13} \\ 
R_{21} & R_{22} & R_{23} \\	
R_{31} & R_{32} & R_{33} \\	
\end{pmatrix}
\times  \begin{pmatrix}
 x \\ 
 y \\	
 z \\	
\end{pmatrix}
\end{equation} 

Each ring, labeled n, may be independent from the others and is inclined with respect to the galaxy reference plane by the angles $\delta_n$ and $\alpha_n$, shown in Figure A.1. The change in $\delta_n$ represent the warping of the gas plane, while the change in $\alpha_n$ is the ring twisting, if present. The coordinate system of each ring is then 
\begin{equation}
\begin{pmatrix}
 x''_n \\ 
 y''_n \\	
 z''_n \\	
\end{pmatrix}
=
 \begin{pmatrix}
 R''_{11n} & R''_{12n} & R''_{13n} \\ 
 R''_{21n} & R''_{22n} & R''_{23n} \\	
 R''_{31n} & R''_{32n} & R''_{33n} \\	
\end{pmatrix}
\times  \begin{pmatrix}
 x_n \\ 
 y_n \\	
 z_n \\	
\end{pmatrix}
\label{matrix}
\end{equation} 

\begin{equation}
\text{with the additonal conditions} 
\begin{cases}
x''=0 & \text{Sky plane} \\
z\  \ =0 & \text{Ring plane} \\
\end{cases} 
\label{planes}
\end{equation}

The values of the coefficients R$_{ijn}(i,\phi,\delta_n,\alpha_n)$ are listed in Arnaboldi \& Galletta (1993) and each ring is described by its radius r$_n$ and by an angle $0^\circ\le\beta\le360^\circ$. 

\begin{equation}
\begin{cases}
x_n= r_n cos\beta_n \\
y_n= r_n sin\beta_n \\
\end{cases}
\label{cilindric}
\end{equation}

To extract a  simulated rotation curve, we calculate the couples of value \{y'',z''\} on the sky traced along a fixed P.A. according to the relation:

\begin{equation}
z''= K y'' 
\end{equation}

where K=tan(90+P.A.). With this assumption, and the definition of the ring plane (equation \ref{planes}) the equation \ref{matrix} becomes:

\begin{equation}
\begin{cases}
y''_n= r (R''_{21n}\ cos\beta_n+R''_{22n}\ sin\beta_n) \\
z''_n= Ky''= r (R''_{31n}\ cos\beta_n+R''_{32n}\ sin\beta_n) \\
\end{cases}
\end{equation}
 
that gives for each \{y'',z''\} point a unique $\beta$ value crossing the n-th ring:

\begin{equation}
tg \beta_n = -\frac{R''_{31n}-K\ R''_{21n}}{R''_{32n}- K\ R''_{22n}}
\end{equation}
 
At this point, the rotational velocity of the n-th ring projected along the line of sight is

\begin{equation}
V_{oss}[y'',z'',n) = V_{circ}(r_n)({R''_{12n}\ cos\beta_n - R''_{11n}}\ sin\beta_n)
\end{equation}

where V$_{rot}$ is the intrinsic circular velocity of the rings. To interpolate the observed rotation values V$_{oss}$ we used a simple rotation curve described by 

\begin{equation}
	V_{circ}(r_n)=3 V_{max}\frac{r_n/r_{max}}{1+2(r_n/r_{max})^{3/2}}
\end{equation}

according to equation [26] and [28] discussed by Brandt (1960).

\end{document}